\documentclass{PoS}

\title{Particle Production in Au+Au Collisions at $\sqrt{s_{NN}}$ = 9.2 GeV}

\ShortTitle{Particle Production in Au+Au Collisions at 9.2 GeV}

\author{\speaker{Jiayun Chen} (for STAR Collaboration)%\thanks{A footnote may follow.}
\\
       Institution of Particle Physics, Central China Normal University (HZNU), Wuhan 430079, P.R.China \\
       Physics Department, Brookhaven National Laboratory, Upton, NY 11973,
       USA\\
       The Key Laboratory of Quark and Lepton Physics (HZNU), Ministry of Education, Wuhan, 430079,
       P.R.China \\
       E-mail: \email{chenjy@iopp.ccnu.edu.cn}}

\abstract{In this report we present the first test run results from
Au+Au collisions at $\sqrt{s_{NN}}$ = 9.2 GeV at RHIC. The large
acceptance STAR detector has collected ~3k minimum bias collisions
during this test run. The azimuthal anisotropy, identified particle
spectra, particle ratios and HBT radii are observed to be consistent
with the previous measurements from CERN SPS at similar center of
mass energies. These results from the lowest collision energy at
RHIC demonstrate the STAR detector's readiness to collect high
quality data for the proposed Critical Point Search Program which
allows us to explore the QCD phase diagram. }

\FullConference{5th International Workshop on Critical Point and Onset of Deconfinement - CPOD 2009,\\
        June 08 - 12 2009\\
        Brookhaven National Laboratory, Long Island, New York, USA}

\begin{document}

\section{Introduction}
In the first decade of STAR running, the evidence about the
existence of a strongly coupled Quark Gluon Plasma (sQGP)
\cite{STARwhitepaper} came in the form of strong suppression of
particle production at large $p_{T}$ \cite{STARhighpT} and the large
amount of elliptic flow \cite{STARFirstFlowPaper}. As an important
step towards understanding the properties of sQGP and the structure
of the QCD phase diagram, a systematic analysis of particle
production as a function of collision energy is necessary.
Theoretical calculations have indicated that the order of the
transition from hadronic matter to the sQGP depends on the baryon
chemical potential ($\mu_{B}$) and temperature ($T$): at low
$\mu_{B}$ and high $T$, a cross-over transition occurs\cite{LargeT};
at high $\mu_{B}$ and low $T$, the phase transition is thought to
become first order \cite{Scavenius:2000qd}; this first order phase
transition ``meets'' with the smooth cross-over at the critical
point\cite{CP}. Experimentally we can access this phase diagram and
vary these initial conditions by changing the beam energy. Thus a
beam energy scan (BES) program will help us to explore the QCD phase
diagram and to locate the critical point \cite{helenPaper,STARnote}.
As a first step of the BES program, RHIC made a test run for Au+Au
collisions at $\sqrt{s_{NN}}$ = 9.2 GeV. The preliminary results
from STAR will be reported in this article.

\section{Experiment and Analysis}
The data presented here are from Au+Au collisions at $\sqrt{s_{NN}}$
= 9.2 GeV, recorded by the STAR experiment at RHIC. The main
detector used to obtain the results on anisotropy flow, yields,
particle ratios and pion interferometry is the Time Projection
Chamber (TPC) \cite{STARTPC}. The TPC is STAR's primary detector and
can track up to 4000 particles per event. For the collisions near
its center, the TPC covers the pesudorapidity ($\eta$) range
$|\eta|\le 1.5$. The full azimuthal coverage of the TPC makes it
ideal for flow and event-by-event fluctuations measurements. The
particle identification (PID) is accomplished by measuring the
ionization energy loss ($dE/dx$) in the TPC. In the future, the Time
of Flight (ToF) detector \cite{STARTOF} (with $2\pi$ azimuthal
coverage and $|\eta |< 1.0$ ) will further enhance the PID
capability. The ToF will be ready for Run10 in December of 2009.

All events were taken with a minimum bias tigger. The tigger
detectors used in this data are the Beam-Beam-Counter (BBC) and
Vertex Position Detector (VPD) \cite{STARtrigger}. The BBCs are
scintillator annuli mounted around the beam pipe beyond the east and
west pole-tips of the STAR magnet at about 375 cm from the center of
the nominal interaction region (IR), and they have the $\eta$
coverage of $3.8<|\eta|<5.2$ and a full azimuthal ($2\pi$) coverage.
The BBCs are used to reconstruct the first order event plane for the
directed flow analysis.

The only result obtained from the forward rapidity windows are the
directed flow measurement, for which we used data taken by the
Forward Time Projection Chamber (FTPC) \cite{STARFTPC}. The FTPC
detects charged particles in $|\eta|$ range from 2.5 to 4.2 with
full azimuthal coverage.  The detail about the design and the other
characteristics of the STAR detector can be found in Ref.
\cite{STARdetector}.

There are $\sim$ 3000 good events collected at about 0.6 Hz in the
year 2008. The data taking period lasted less than 5 hours.

The centrality definition and the tracks selection for various
analysis are explained in Ref. \cite{Lokesh}.

\section{Results and Discussion}
\begin{figure}
%\resizebox{0.45\textwidth}{!}
{\includegraphics[width=0.5\textwidth,height=2.45in]{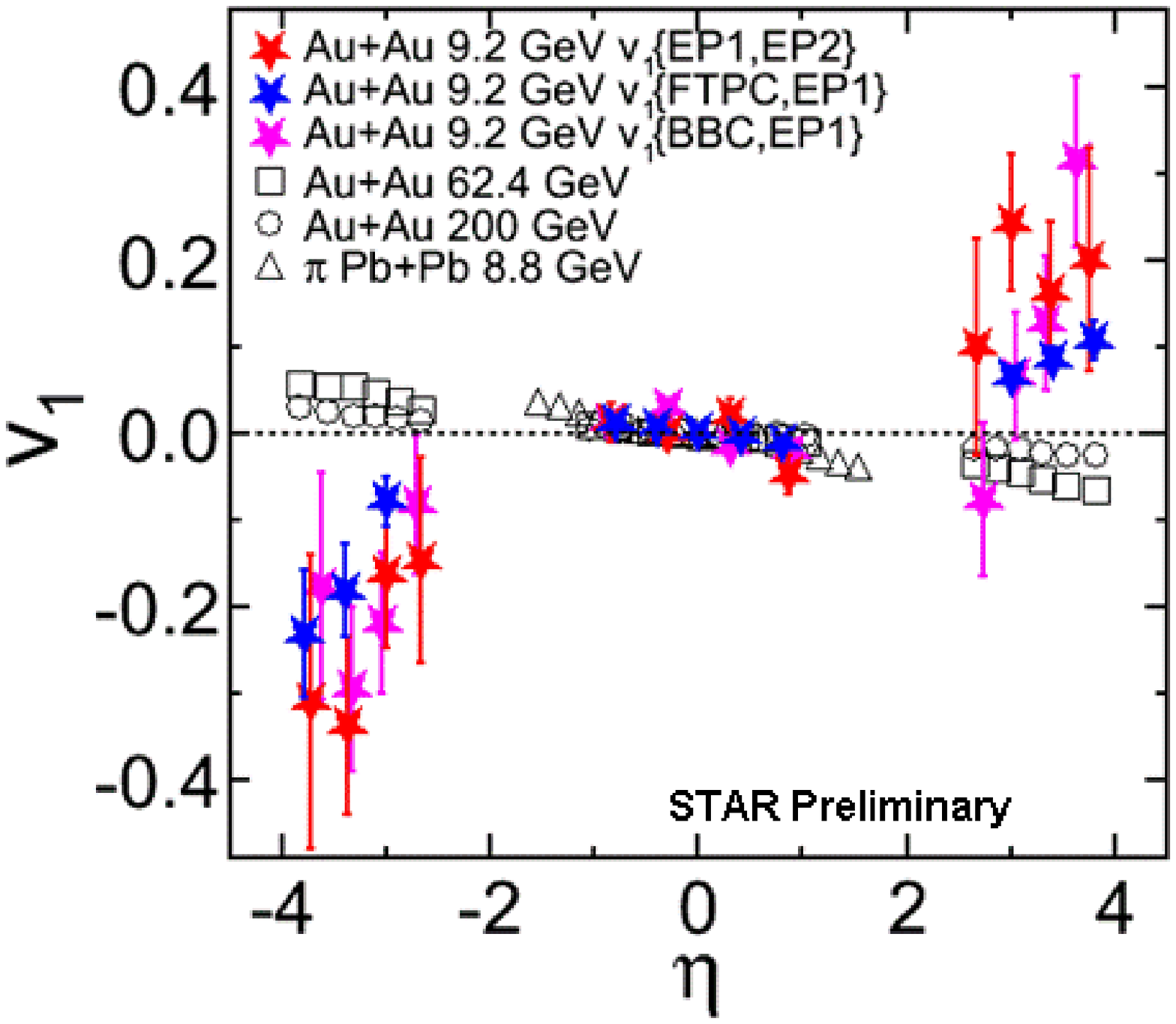}}
%\resizebox{0.43\textwidth}{!}
{\includegraphics[width=0.5\textwidth,height=2.45in]{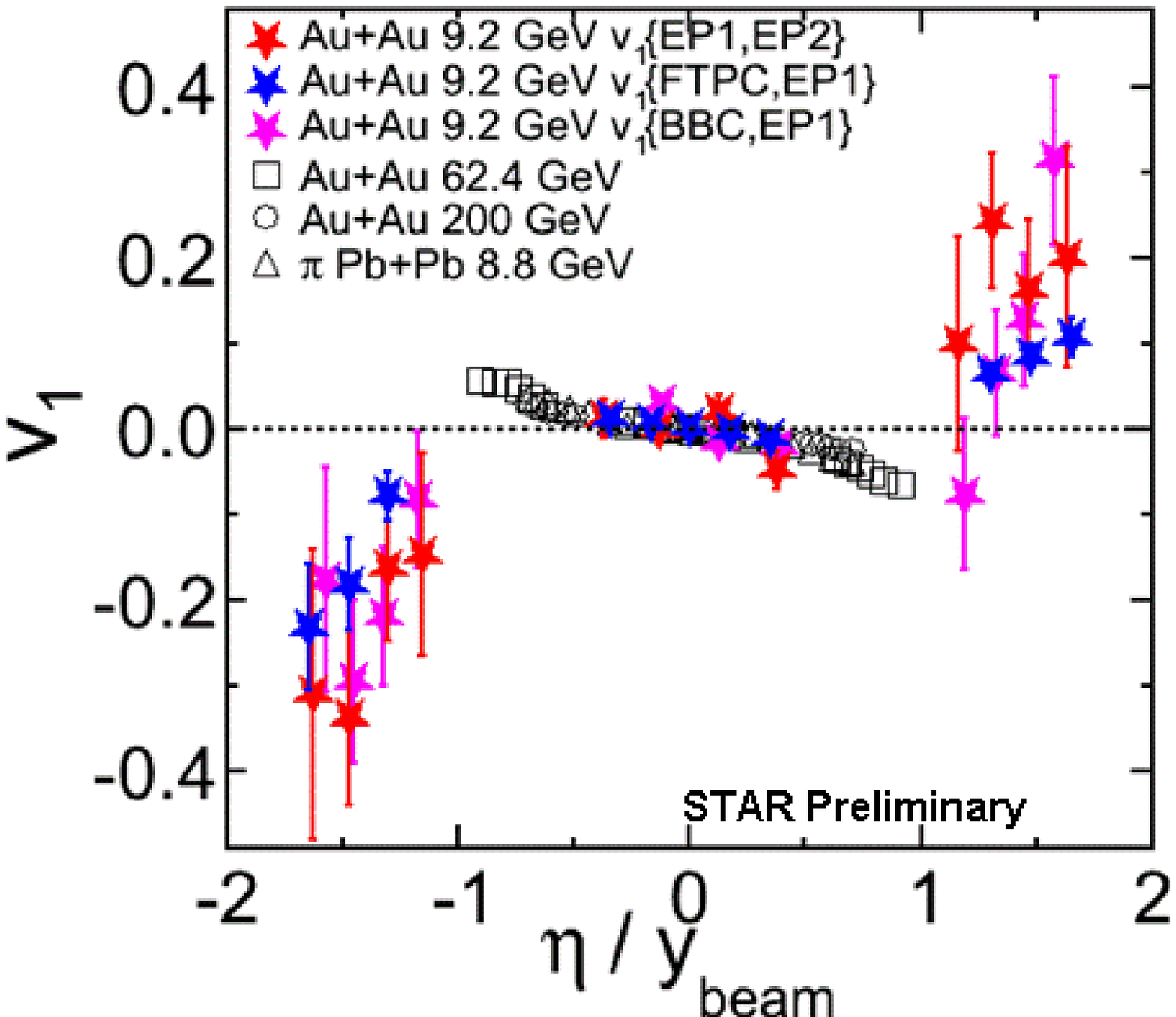}}
\caption{Left panel: Charge hadron $v_{1} vs. \eta$ from Au+Au
collisions at 9.2 GeV in centrality 0-60\%. The solid star symbols
are the results obtained from the mixed harmonic method (red), the
standard methods which include the first order event plane
reconstructed by FTPC (blue) and BBC (pink) respectively (see text
for details). The results are compared with $v_{1}$ from Au+Au
collisions at 62.4 GeV (square symbols) and 200 GeV (circle symbols)
from centrality 30-60\% \cite{starv1paper}. The charged pions
$v_{1}$ from Pb+Pb collisions at 8.8 GeV are also shown
\cite{NA49v1}. Right panel: Same as the left panel, but plotted as a
function of $\eta/y_{beam}$. The error bars include only statistical
uncertainties.} \label{fig1}
\end{figure}

Fig. 1 (left panel) shows the charged hadrons $v_{1}$ as a function
of $\eta$ in 0-60\% most central Au+Au collisions at 9.2 GeV.  The
$p_{T}$ range for this study is 0.15-2.0 GeV/c. The $v_{1}$ results
from 9.2 GeV are obtained by different methods: 1) The standard
methods: the one for which the first order event plane is
reconstructed from the FTPC tracks is named $v_{1}\{EP_{1},FTPC\}$,
while that which uses BBC hits for the first order event plane
reconstruction is named $v_{1}\{EP_{1},BBC\}$; 2) The mixed
harmonics method, denoted by $v_{1}\{EP_{1},EP_{2}\}$. This method
utilizes the large elliptic flow signal at RHIC, and at the same
time suppresses the non-flow arising from the correlation of
particles from the same harmonics. The details of these methods can
be found in Ref. \cite{mixedHar}. The $v_{1}$ results at 9.2 GeV
from different methods are consistent within the error bars. These
results of $v_{1}$ from $\sqrt{s_{NN}}$ = 9.2 GeV Au+Au collisions
are compared to corresponding results in 30\%-60\% central Au+Au
collisions at 62.4 and 200 GeV \cite{starv1paper}.
%These results are
%also compared with charged pion's $v_{1}$ in Pb+Pb collision at
%$\sqrt{s_{NN}}$ = 8.8 GeV. At mid-rapidity, all the results have
%comparable values. At forward rapidities, the trend of $v_{1}$ from
%low energy is different from that at higher energies. The rapidity
%distribution of the $v_{1}$ depends on the early longitudinal
%collision dynamics and interactions between the spectators and
%particles originating from the overlapped region. However, as one
%can see in Fig. 1 (right panel), when we plot $v_{1}$ as a function
%of the normalized rapidity, $\eta/y_{beam}$, all values of $v_{1}$
%seems lie on a common trend. Here the term $y_{beam}$ is the
%corresponding nucleon beam rapidity for a given beam energy.
The results are compared with $v_{1}$ of charged pions in Pb+Pb
collisions at $\sqrt{s_{NN}}$ = 8.8 GeV. At and near mid-rapidity,
the $v_{1}$ values are independendent of beam energy. At forward
rapidities, the $v_{1}$ values seem to change sign at lower
colliding energy while those at higher energies retain the monotonic
behaviour. However, if $v_{1}$ are plotted as a function of $\eta$
scaled with the beam rapidity $y_{beam}$, as shown in the right plot
of Fig. 1, in the common region of values of $\eta/y_{beam}$ the
$v_{1}$ values remain the same at all energies.

\begin{figure}
%\resizebox{0.45\textwidth}{!}
{\includegraphics[width=0.5\textwidth,height=2.45in]{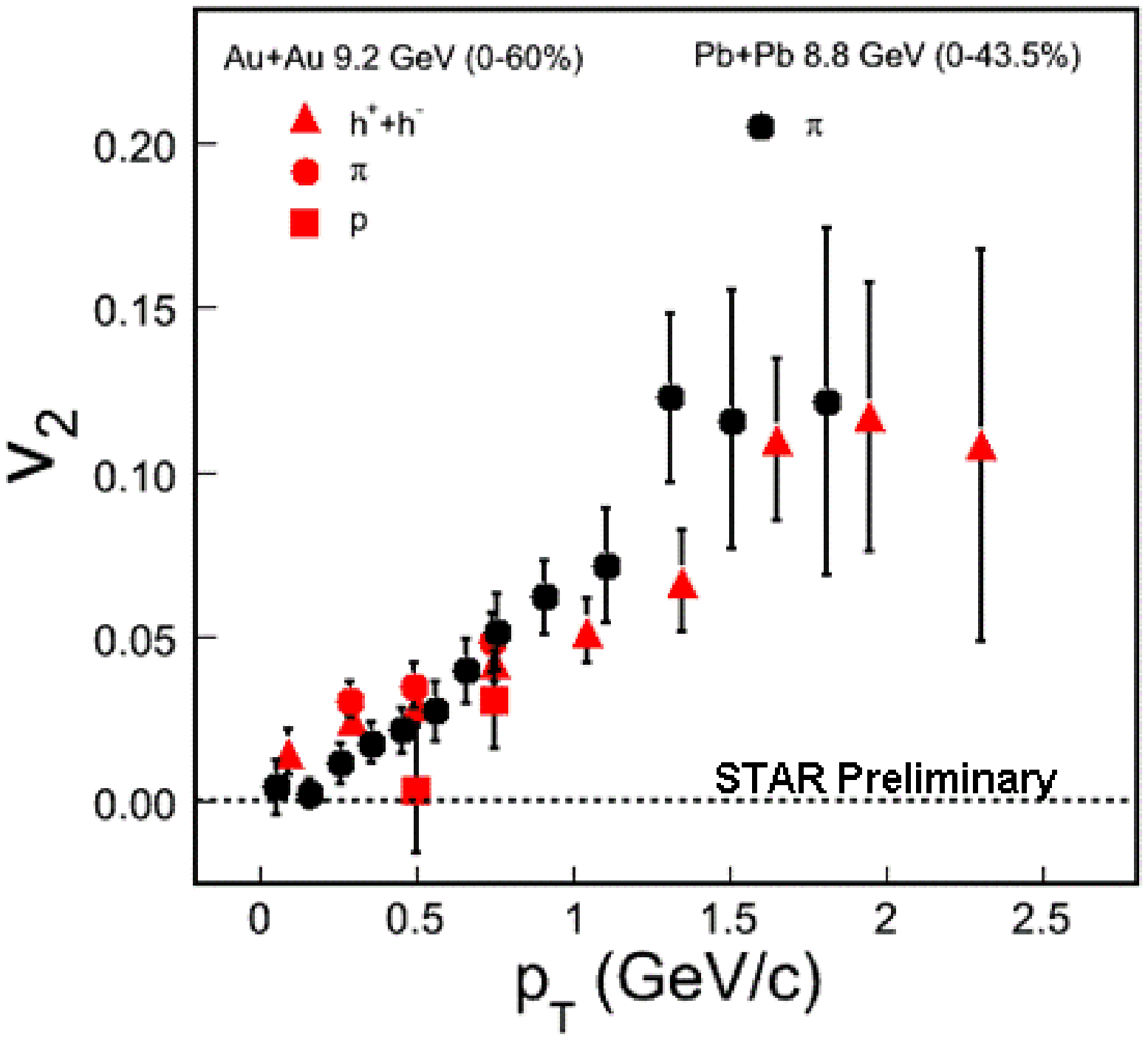}}
%\resizebox{0.43\textwidth}{!}
{\includegraphics[width=0.5\textwidth,height=2.45in]{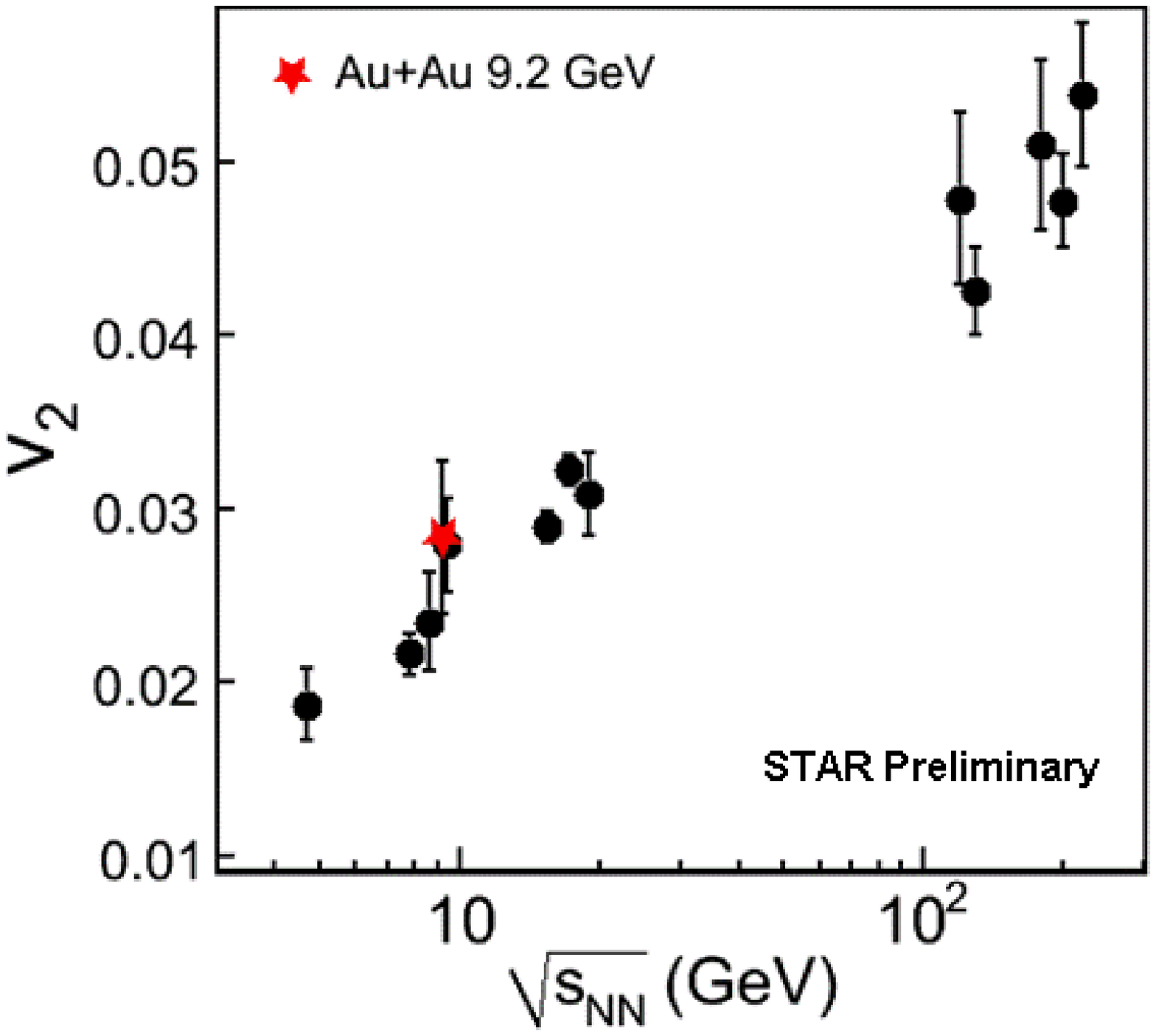}}
\caption{Left panel: $v_{2}$ as function of $p_{T}$ for charged
hadrons (red triangles), pions (red circles) and protons (red
squares) in 0-60\% Au+Au collisions at 9.2 GeV. For comparison,
$v_{2}(p_{T})$ results for pions(black circles) from NA49
\cite{NA49v1} in 0-43.5\% Pb+Pb collisions at 8.8 GeV are shown.
Right panel: Energy dependence of integral $v_{2}$ ($|\eta|<1.0$)
for Au+Au collisions at 9.2 GeV from centrality 0-60\%. The STAR
charged hadron $v_{2}$ \cite{STARv2paper} results are compared with
the measurement from E877 \cite{E877v2paper}, NA49 \cite{NA49v1},
PHENIX \cite{PHENIXv2paper}, and PHOBOS
\cite{PHOBOS1,PHOBOS2,PHOBOS3}. Only statistical error is
shown.}\label{fig2}
\end{figure}

The elliptic flow versus transverse momentum for the charged
hadrons, pions and protons from the 9.2 GeV Au+Au collisions are
shown in Fig. 2 (left panel). The comparison with the similar energy
from NA49  \cite{NA49v1} are also shown. The STAR and NA49 (at
$\sqrt{s_{NN}}$ = 8.8 GeV) results seem to be consistent within
errors. Fig. 2 (right panel) shows intergral $v_{2}$ as a function
of beam energy for charged hadrons. Results from minimum bias
collisions at 9.2 GeV at midrapidity are compared with those from
STAR at higher energy \cite{STARv2paper} and other experiments
\cite{NA49v1,E877v2paper,PHENIXv2paper,PHOBOS1,PHOBOS2,PHOBOS3}. The
$v_{2}$ results nicely fits into the observed trends.

\begin{figure}
\centering
{\includegraphics[width=0.7\textwidth]{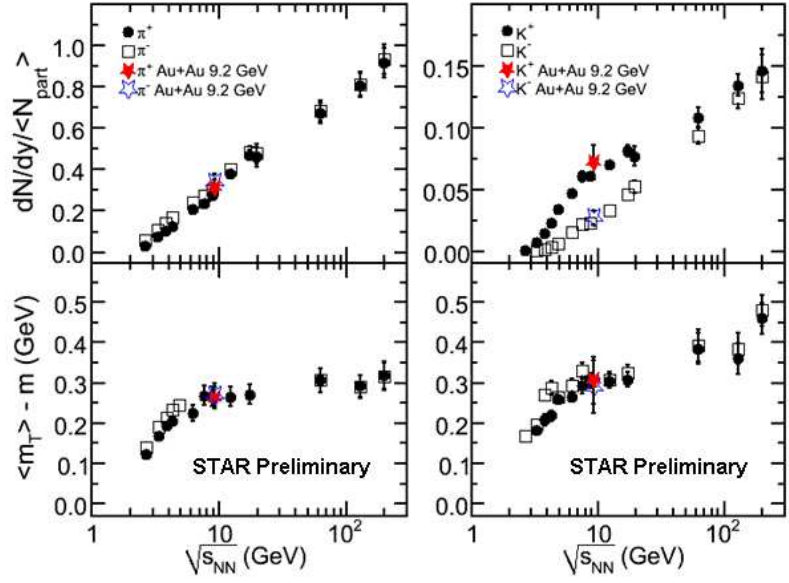}}
\caption{$dN/dy$ normalized by $\langle N_{part}\rangle$ (upper
panels) and $\langle m_{T}\rangle-m$ (lower panels) of $\pi^{\pm}$
(left panels) and $K^{\pm}$ (right panels) in 0-10\% central Au+Au
collisions for 9.2 GeV compared to previous results from AGS
\cite{AGSyield},SPS \cite{SPSyield} and RHIC \cite{STARyield}. The
errors are statistical and systematic errors added in quadrature.}
\label{fig3}
\end{figure}

Fig. 3 (upper panels) shows the particle yields per unit rapidity
$dN/dy$ from $\pi^{\pm}$ and $K^{\pm}$ as a function of
$\sqrt{s_{NN}}$ at midrapidity for central collisions. Fig 3 (lower
panels) shows the $\langle m_{T}\rangle -m_{0}$ for $\pi^{\pm}$ and
$K^{\pm}$ respectively in 0-10\% central Au+Au collisions at 9.2
GeV. These results are compared to those at AGS, SPS and RHIC
energies \cite{AGSyield,SPSyield,STARyield}. The yield and $\langle
m_{T}\rangle-m_{0}$ beam energy dependence from 9.2 GeV are
consistent with the published data. The $\langle m_{T}\rangle-m_{0}$
seems to increase with the energy at the AGS energy region
$\sqrt{s_{NN}} \sim$ 2-5 GeV. It appears to be independent of the
energy in the SPS energy window, $\sqrt{s_{NN}} \sim$ 5-20 GeV. As
shown in the figure, the RHIC 9.2 GeV collision data is similar to
other results. Assuming a thermodynamic system, the $\langle
m_{T}\rangle-m_{0}$ is associated with temperature and the $dN/dy
\varpropto \log (\sqrt{s_{NN}})$ represents the entropy of the
system. It has been argued that the observations in this scenario
could reflect the characteristic signature of a first order phase
transition \cite{VanHove}. The energy independence of $\langle
m_{T}\rangle-m_{0}$ , $\sqrt{s_{NN}} \sim$ 8-60 GeV, can be
interpreted as reflecting of a mixed phase.

\begin{figure}
\centering
{\includegraphics[width=0.7\textwidth]{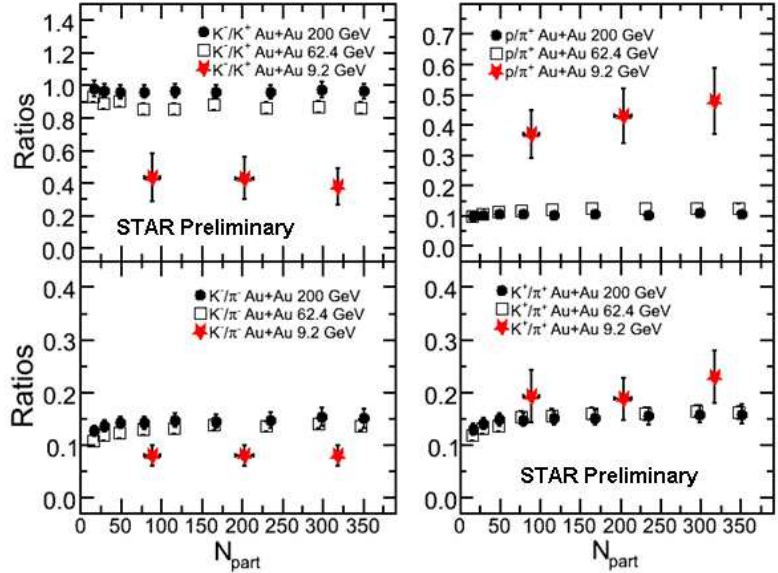}}
\caption{Variation of $K^{-}/K^{+}$ (top left panel),
$K^{-}/\pi^{-}$ (bottom left panel), $p/\pi^{+}$ (top right panel)
and $K^{+}/\pi^{+}$ (bottom right panel) from Au+Au collision at 9.2
GeV. The corresponding results from Au+Au collisions at 62.4 GeV and
200 GeV \cite{STARyield,STARratio} are also shown for comparison.
The errors shown are systematic and statistical errors added in
quadrature.} \label{fig4}
\end{figure}

Fig. 4 shows the centrality dependence of particle
ratios($K^{-}/K^{+}, K^{-}/\pi^{-}, p/\pi^{+}$ and $K^{+}/\pi^{+}$)
from 9.2 GeV (red star symbols) compared with STAR results at high
energies 62.4 GeV (open square symbols) and 200 GeV (full circle
symbols) \cite{STARyield,STARratio}. The $K^{-}/K^{+}$ and
$K^{-}/\pi^{-}$ ratios are lower in 9.2 GeV compared to that in 200
and 62.4 GeV. On the other hand, %the $K^{+}/\pi^{+}$ ratio in 9.2
%GeV is comparable to those at 200 and 62.4 GeV. This reflects that
%the $K^{+}$ production is dominated by associated production in low energy.
There is less variation between 9.2 GeV and the highest RHIC
energies in the $K^{+}/\pi^{+}$ ratio than in case of the other
particle ratios just discussed. This reflects that the production of
kaons at the mid-rapidity going from a region of high net-baryon
density at lower collision energies, to a region of lower net-baryon
density at high beam energies. $K^{+}$ production at lower beam
energies is dominated by associated production. The $p/\pi^{+}$
ratio in 9.2 GeV is greater compared to that in 200 GeV and 62.4
GeV, reflecting a large baryon stopping at the low energy. All these
particle ratios presented seem to be independent of collision
centrality for all the energies studied.

\begin{figure}
\centering
{\includegraphics[width=0.7\textwidth]{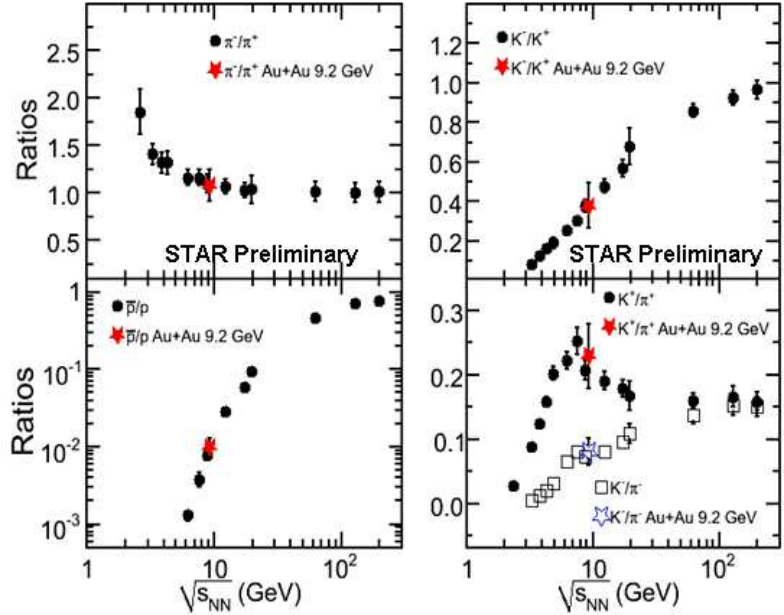}}
\caption{The particle ratios at $|y|<0.5$ for  0-10\% central 9.2
GeV Au+Au collisions are compared to previous results from AGS
\cite{AGSyield}, SPS \cite{SPSyield} and RHIC \cite{STARyield}
:$\pi^{-}/\pi^{+}$(top left panel), $\bar{p}/p$(bottom left panel),
$K^{-}/K^{+}$(top right panel) and $K/\pi$(bottom right panel). The
errors shown are statistical and systematic errors added in
quadrature.}\label{fig5}
\end{figure}

Fig. 5 shows the collision energy dependence of particle ratios
($\pi^{-}/\pi^{+}, K^{-}/K^{+}, \bar{p}/p$ and $K/\pi$). The results
from Au+Au 9.2 GeV (red star symbol) seem to follow the trends
established by the previous measurements
\cite{AGSyield,SPSyield,STARyield}. In Au+Au 9.2 GeV collisions, the
$\pi^{-}/\pi^{+}$ ratio is about 1 which shows that this production
is dominated by pair production. The $K^{-}/K^{+}$ ratio is around
0.4, indicating the significant contribution from the associated
production for the positive kaons. The $\bar{p}/p$ ratio is far less
than one which reflect large baryon stopping. The $K/\pi$ ratio is
of special interest as it expresses the enhancement of strangeness
production relative to non-strange hadrons. The $K^{+}/\pi^{+}$
ratio increases as a function of beam energy and reaches a peak at
about $\sqrt{s_{NN}} \sim$ 7.7 GeV. At higher beam energies, the
ratio starts to show  a decreasing trend. The new 9.2 GeV results
are found to be consistent with previous observed energy dependence
within the error bars. Note that only the ratio of positive
particles, $K^{+}/\pi^{+}$, shows the peak at about 8 GeV. For the
negative ratios, a monotonic increase as a function of beam energy
is seen, see open-squares in the figure. Since only the $K^{+}$ is
involved in the associate process ($N+N \rightarrow N+
\Lambda+K^{+})$, the observed peak may reflect the maximum stopping
reached in high-energy nuclear collisions. The RHIC BES program will
make precision measurements for all of these ratios at this
interesting energy region in the near future.

\begin{figure}
\centering {\includegraphics[width=0.6\textwidth]{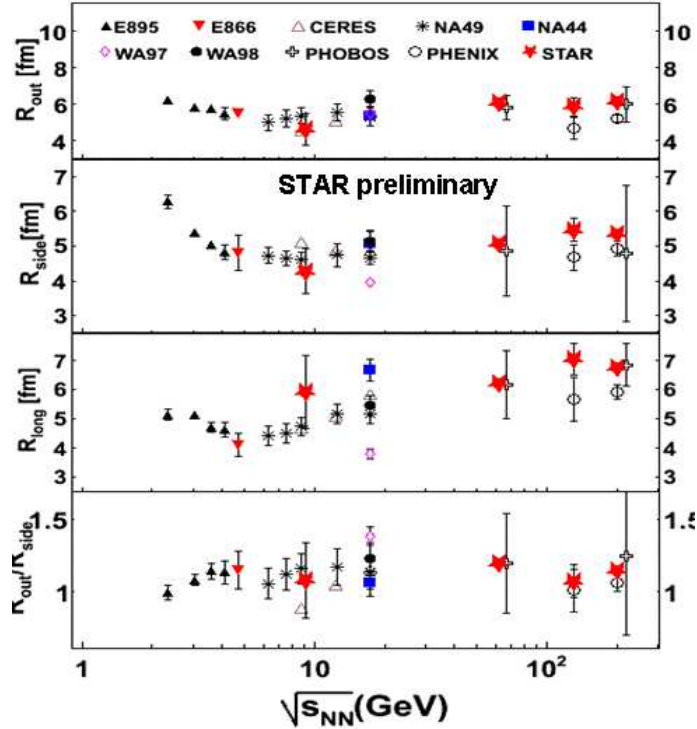}}
\caption{The HBT radii $R_{out},R_{side},R_{long}$ and the ratio
$R_{out}/R_{side}$ for $\pi^{-}$ are plotted as a function of beam
energy. The centrality is 0-60\% and the $k_{T}$ range is
$[0.15,0.25]$ (GeV/c) for Au+Au 9.2 GeV. The errors for Au+Au 9.2
GeV are only statistical. The systematic errors are expected to be
less than 10\% for all radii. The 9.2 GeV results are compared with
the measurements by STAR high energies \cite{STARHBT}, PHENIX
\cite{PHENIXHBT}, PHOBOS \cite{PHOBOSHBT}, NA44 \cite{NA44HBT}, NA49
\cite{NA49HBT}, E802 \cite{E802HBT}, E866 \cite{E866HBT}, E895
\cite{E895HBT}, WA97 \cite{WA97HBT}, CERES
\cite{CERESHBT}.}\label{fig6}
\end{figure}

The $\pi$ interferometry from low $p_{T}$ and mid-rapidity give us
information about the space-time size of emitting source and
freeze-out process of the fireball \cite{STARHBT}. Fig. 6 shows the
results for Hanbury-Brown-Twiss (HBT) pion ($\pi^{+}$)
interferometry measurements. There are three radii parameters
$R_{out},R_{side}$ and $R_{long}$ measured in our analysis. The
$R_{out}$ measures the spatial and temporal extension of the source
whereas the $R_{side}$ gives the spatial extension of the source.
Thus the ratio $R_{out}/R_{side}$ reflects the emission duration of
the source. It is expected that the ratio $R_{out}/R_{side}$ is much
larger than 1 for a first order transition \cite{HBT}. For the
measured beam energies
\cite{STARHBT,PHENIXHBT,PHOBOSHBT,NA44HBT,NA49HBT,E802HBT,E866HBT,E895HBT,WA97HBT,
CERESHBT}, the ratio $R_{out}/R_{side}$ is close to 1 within errors.

\section{Summary and Outlook}
In summary, the azimuthal anisotropy ($v_{1}$ and $v_{2}$) results
from Au+Au 9.2 GeV are similar to those obtained from collisions at
similar energies. The identified particle spectra are obtained from
Au+Au collisions at 9.2 GeV. The centrality and beam energy
dependence of the hadron yields and ratios are studied. In the
central collisions, the anti-proton to proton ratio $\sim$ 0.01
shows significant baryon stopping in these collisions. The
$K^{-}/K^{+}$ ratio from the central collision is $\sim$ 0.4,
showing associated production for $K^{+}$. The $p/\pi$ ratio is
higher and $K^{-}/\pi^{-}$ is lower at 9.2 GeV compared to 200 GeV
at all collision centralities studied. The pion interferometry
results in the Au+Au collisions at 9.2 GeV follow the established
beam energy trends.

These 9.2 GeV results from Au+Au collision are the first important
step towards the RHIC BES program. There are several key
measurements for the program: 1) The PID hadron spectra, ratios and
elliptic flow give the information about the collectivity dynamics;
2) The fluctuations including net-proton kurtosis, $K/\pi$ ratio,
$\langle p_{T}\rangle , \langle N\rangle$ access the thermodynamics
of the system. Our BES program is planned to cover the energy range
$\sqrt{s_{NN}} \sim 5-39$ GeV.
%All the theoretical estimations of the Critical Point are within our BES search area.
Several Lattice QCD based estimates of the Critical Point
\cite{LargeT,Scavenius:2000qd,CP} are within our BES search area.
The results presented in this proceedings from the lowest beam
energy collisions at RHIC demonstrate the STAR detector's capability
to collect high quality data for the proposed BES Program. Large and
uniform acceptance for all beam energies in a collider set-up,
excellent particle identification (TPC+TOF) and higher statistics
will lead to significant advances in the Critical Point search.

\end{document}